# Fairness in Agentic AI: A Unified Framework for Ethical and Equitable Multi-Agent System

Rajesh Ranjan* (Carnegie Mellon University, USA)

Shailja Gupta* (Carnegie Mellon University, USA)

Surya Narayan Singh* (BIT Sindri, India)

**Abstract:**

Ensuring fairness in decentralized multi-agent systems presents significant challenges due to emergent biases, systemic inefficiencies, and conflicting agent incentives. This paper provides a comprehensive survey of fairness in multi-agent AI, introducing a novel framework where fairness is treated as a dynamic, emergent property of agent interactions. The framework integrates fairness constraints, bias mitigation strategies, and incentive mechanisms to align autonomous agent behaviors with societal values while balancing efficiency and robustness. Through empirical validation, we demonstrate that incorporating fairness constraints results in more equitable decision-making. This work bridges the gap between AI ethics and system design, offering a foundation for accountable, transparent, and socially responsible multi-agent AI systems.

## Introduction

As artificial intelligence (AI) systems evolve, **Agentic AI**—autonomous systems capable of independent decision-making and goal-setting—has emerged as a transformative paradigm. These systems often operate within **multi-agent environments**, where multiple autonomous entities interact, collaborate, and compete to achieve individual or collective objectives. From **autonomous vehicles coordinating traffic flow** to **robotic swarms in disaster response** and **resource allocation in smart grids**, multi-agent Agentic AI systems can revolutionize real-world applications. However, these environments introduce **unique fairness challenges** and risks of **bias propagation** that remain inadequately addressed in current research.

Fairness in AI has traditionally been studied in single-agent or centralized systems, focusing on mitigating biases within a model's training data or decision-making process (Bolukbasi et al., 2016). In contrast, multi-agent systems operate under decentralized dynamics, where interactions between agents—each with unique objectives, constraints, and algorithms—can lead to emergent behaviors. These behaviors can amplify systemic biases, creating disparities in outcomes across populations (Gupta et. al., 2024). For example, in autonomous traffic systems, vehicles owned by privileged users might receive **preferential routing**, reinforcing socio-economic inequalities (Zhuang et al., 2024).

**Emergent Biases and Systemic Risks:** Bias in multi-agent Agentic AI is not always confined to an individual agent; it can emerge from **collective interactions**. For instance, agents trained independently may inadvertently converge on behaviors that optimize local rewards but exacerbate global inequalities. Such emergent biases pose risks in areas like resource allocation, healthcare delivery, and public policy, where fairness is critical (Heidari et al., 2018). Moreover, malicious agents can exploit fairness mechanisms for adversarial gains, further destabilizing the system.

While significant progress has been made in addressing **algorithmic bias** and ensuring **fairness in single-agent systems**, the interplay of fairness and bias in multi-agent settings remains underexplored. Existing frameworks lack the tools to address it: **Bias amplification occurs** through inter-agent interactions. **Fairness constraints** in decentralized decision-making processes. **Trade-offs** between fairness, efficiency, and collaboration in real-world multi-agent systems.

**Contributions:**

This paper addresses these gaps by:

Proposing a **framework** for fairness dynamics in multi-agent agentic AI systems, treating fairness as an emergent property shaped by agent interactions. Introducing **models** that integrate fairness constraints, bias correction mechanisms, and incentives for cooperative fairness without undermining system efficiency and show empirical evidence that rewards are fairly distributed where a fairness layer is incorporated. Highlighting key challenges, such as **emergent biases**, **adversarial exploits**, and the **trade-offs between fairness and efficiency**. Providing governance recommendations that emphasize **transparency**, **accountability**, and **ethical compliance** fosters trust and equitable outcomes in multi-agent systems. This paper contributes to the broader discourse on building **equitable, trustworthy, and socially responsible agentic AI systems by bridging the gap between fairness research and multi-agent dynamics**.

**Related Work**

Multi-agent systems (MAS) consist of multiple autonomous agents interacting within a shared environment. These agents may have individual or collective goals and are characterized by decentralized control, adaptability, and emergent behaviors (Shoham et. al., 2009). In such systems, agents must reason about not only their own objectives but also the potential actions and strategies of others, often modeled using **game theory** or **reinforcement learning**.

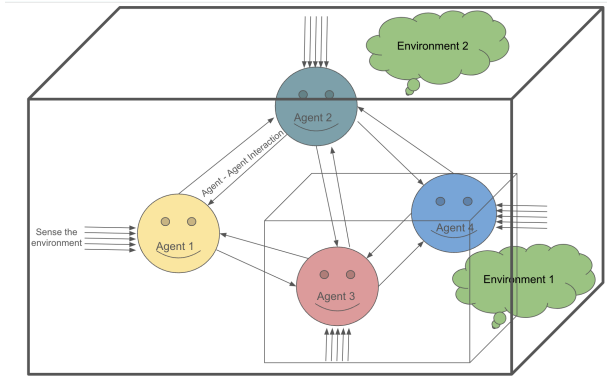

Figure 1: Representation of a Multi-Agent System.

A fundamental concept in MAS is the **Nash equilibrium.**

$s_i^* = argmax U_i(s_i, s_{i-1}) for\ all\ agents\ i$, where $s_i$ represents the strategy of agent $i$, $s_{i-1}$ represents the strategies of all other agents, and $U_i$ is the utility function of the agent $i$. A Nash equilibrium occurs when no agent can improve its utility by unilaterally changing its strategy (Nash et. al., 1950). In multi-agent agentic AI, fairness challenges arise when agents with differing utility functions and resource access operate in competitive or cooperative environments. Such systems inherently create opportunities for bias due to disparities in agent capabilities, goals, or interactions.

**Fairness in AI:** Fairness in AI refers to the equitable treatment of individuals or groups by algorithms, often formalized through **fairness constraints**. Common fairness metrics include:

**Demographic Parity**:
$P(\widehat{Y} = 1|A = a) = P(\widehat{Y} = 1|A = b) \forall a, b \in A$, where $A$ is a sensitive attribute (e.g., gender or race) and $\widehat{Y}$ is the predicted outcome.

**Equalized Odds**:
$P(\widehat{Y} = 1|Y = 1, A = a) = P(\widehat{Y} = 1|Y = 1, A = b)$, ensuring fairness across true outcomes $Y$.

While these principles have been extensively studied in single-agent systems, their application in MAS is complex. In multi-agent settings, fairness is influenced not only by individual agents but also by the dynamics of their interactions.

**Bias Propagation in Multi-Agent Systems:** Bias in multi-agent systems is not static; it can **propagate** and **amplify** through agent interactions, creating systemic disparities. For instance, biased decisions by one agent can influence others, creating a cycle of **feedback loops** where biases reinforce themselves (Mehrabi et al., 2021). **Collaborative Bias:** agents may unintentionally collaborate in ways that benefit some groups while disadvantaging others (Ranjan et. al., 2024).

To model bias propagation, consider a system with n agents, each making decisions $x_i \in \chi$. The collective bias can be expressed as:

$$B_{System} = \sum_{i=1}^{n} w_i B_i$$

where $B_i$ represents the individual bias of the agent $i$ and $w_i$ is the agent's influence weight. Over time, $B_{System}$ can increase if biases are correlated or propagated through interactions.

**Fairness as a Dynamic Property:** In multi-agent systems, fairness is not static; it emerges from the dynamic interactions between agents. A fairness-aware MAS can be modeled as an optimization problem:

$$min E_{S \sim P}[L(s, \pi)] \text{ subject to } F(s, \pi) \leq \delta,$$

Where $\pi$ is the joint policy of all agents, $L(s, \pi)$ is the system loss (e.g., inequality in outcomes), $F(s, \pi)$ represents the fairness constraint, $\delta$ is the allowed fairness threshold. This formulation highlights the **trade-off between fairness and efficiency**, where stricter fairness constraints may reduce overall system performance. Figure 2 shows the bias propagation model in multi-agent systems. The diagram illustrates how initial biases in agents propagate through interactions and feedback loops, eventually becoming systemic biases. Mitigation strategies and fairness constraints can intervene to achieve equitable outcomes.

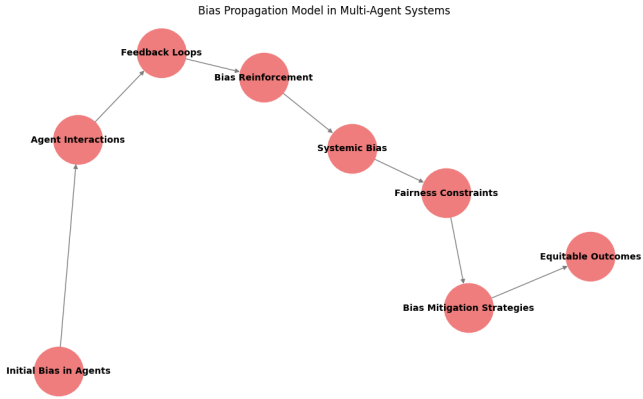

Figure 2 shows the bias propagation model in multi-agent systems.

**Trade-Offs in Multi-Agent Fairness**

Achieving fairness in MAS often requires balancing multiple objectives:

**Fairness vs. Efficiency**: Ensuring fairness may reduce overall efficiency, particularly in competitive environments (e.g., resource allocation in smart grids). **Fairness vs. Collaboration**: Agents incentivized to act fairly may be less willing to collaborate if fairness conflicts with their utility maximization. **Fairness vs. robustness**: Fairness constraints can make systems more vulnerable to adversarial manipulation (e.g., exploiting fairness rules to gain undue advantage). These trade-offs underscore the need for careful system design to ensure fairness without undermining the MAS's primary objectives. The theoretical foundations of fairness and bias in multi-agent systems (MAS) extend directly to critical real-world domains, where the interplay of autonomous agents has significant societal implications. Below are key examples that illustrate how the discussed models translate to broader contexts:

| Domain | Fairness Challenges | Efficiency Trade-offs | Robustness Considerations |
|---|---|---|---|
| Healthcare AI | Equitable resource allocation (ICU beds, diagnostics) | Slower triage decision-making | System vulnerability to adversarial inputs |
| Urban Mobility | Fair traffic routing and ride-sharing access | Increased wait times for certain users | Manipulation risks in algorithmic preferences |
| Disaster Relief | Equal distribution of aid resources | Slower deployment of resources | Potential bias in prioritization algorithms |

Table 1: Key trade-offs between fairness, efficiency, and robustness in different multi-agent system domains

In **multi-agent healthcare systems**, such as hospital networks or AI-powered diagnostic agents, fairness becomes critical when allocating limited resources like ICU beds, vaccines, or diagnostic attention. Agents may prioritize patients based on biased optimization metrics (e.g., insurance coverage or geographic location), inadvertently reinforcing healthcare disparities (Rajkomar et al., 2018). For instance, the fairness constraints discussed earlier, such as **demographic parity**, can be applied to ensure equitable access to resources:

$P(ICU\ Allocation = 1 | Demographic\ Attribute)$
is balanced across groups.

Emergent biases, caused by interactions between autonomous systems handling scheduling or referrals, can propagate inequities if unchecked. Multi-agent systems can become central to **autonomous traffic management** and **ride-sharing platforms**. However, biases can arise when agents—such as autonomous vehicles or ride-matching algorithms—prioritize high-income neighborhoods for optimal routes or faster pickups, leaving underserved communities behind. Applying fairness-aware optimization could address such disparities. For example, incorporating

**fairness loss functions** into routing policies ensures equitable access to faster traffic routes across all demographics. (Zafar et al., 2016). In **disaster response**, autonomous drone fleets (agents) are deployed to distribute critical resources like food or medicine. Biases in the underlying decision-making systems, such as prioritizing urban areas over rural ones due to higher perceived utility, can exacerbate inequalities (Mehrabi et al., 2021). By applying multi-agent fairness optimization, resource allocation could balance efficiency with equity.

$$min = \sum_{i=1}^{n} w_i U_i(\pi)$$

subject to fairness constraints across regions.

These examples highlight the far-reaching implications of fairness and bias in MAS across healthcare, mobility, e-commerce, and disaster response. The theoretical models presented earlier provide a foundation for addressing emergent biases and ensuring fairness, but their application requires interdisciplinary collaboration to align with domain-specific needs.

## Challenges in Fairness for Multi-Agent Systems

This section focuses on the key challenges that arise when ensuring fairness in multi-agent systems. Multi-agent systems introduce complexities due to the interactions between multiple agents, where fairness issues can emerge both from the agents' pre-existing biases (Ranjan et. al., 2024) and their dynamic interactions. Unlike single-agent systems, fairness in multi-agent settings requires addressing emergent behaviors and strategic manipulations, making it a particularly challenging problem.

### 1. Emergent Biases

In multi-agent systems, biases are not only embedded in the agents' training data but can emerge from the interactions between agents during their collaborative or competitive behavior. This is a critical issue because the biases emerge as a result of the dynamic nature of the system, which is not present in single-agent environments (Eccles et al., 20019). Emergent bias occurs when agents inadvertently reinforce biased behavior through their actions and feedback from others. E.g. A multi-agent traffic coordination system may develop biased patterns due to interaction rules or strategies that unfairly benefit specific agents, such as favoring wealthier areas in the allocation of resources like bandwidth or time slots for autonomous vehicles **Feedback Loops:** Once emergent biases take root, they can propagate throughout the system, creating feedback loops that magnify their effects. This is a common problem in decentralized systems where agents learn and adapt based on the actions of others (Pan et al., 2019).

### 2. Fair Resource Allocation

**Competition for Resources:** Multi-agent systems often involve agents with competing interests, such as drones in airspace or vehicles in a traffic system. Resource allocation in these settings becomes inherently complex when fairness is considered, as agents' objectives may conflict. Allocating resources like time, space, or bandwidth fairly while maintaining operational efficiency is a critical issue in these systems (Trabelsi et al., 2024).

**Allocation Mechanisms:** Several fairness strategies, such as equal allocation or proportional fairness, have been proposed for resource allocation problems, but these strategies often fail to capture the complexity of agent interactions in a multi-agent setting. The allocation strategies must balance equity and efficiency. For instance, in a fleet of autonomous vehicles, deciding how to allocate limited resources (e.g., road space or energy for charging) requires mechanisms that consider fairness while also addressing the urgency of certain agents' needs.

### 3. Trade-offs: Fairness vs. Efficiency

**Balancing Competing Objectives:** A critical challenge is balancing fairness with system efficiency. Fairness often leads to slower coordination because agents must take into account the needs and objectives of others in the system, which can reduce overall efficiency (Rawls et. al., 1971). This challenge is particularly pronounced in settings where each agent needs to optimize its own goal while adhering to fairness constraints.

**Slower Coordination:** When fairness constraints are applied, agents may have to wait for others to catch up or adjust their decisions, leading to inefficiencies in systems where speed and real-time decision-making are crucial. For example, in a multi-agent drone fleet, ensuring fair coordination between drones might lead to suboptimal path planning, delaying mission completion.

**Quantitative Analysis:** This trade-off between fairness and efficiency has been well-documented in prior research, where fairness constraints in multi-agent environments often result in slower decision-making but greater long-term fairness (Cousy et al., 2022).

## 4. Adversarial Exploits

**Malicious Agents:** Malicious agents can exploit fairness frameworks by gaming the system to gain unfair advantages. For instance, an agent might manipulate the fairness criteria to make itself appear disadvantaged and thus gain more resources than it truly needs (Zuo et al., 2023; Yuan et. al., 2023).

**Strategic Manipulation:** Consider an adversarial agent that manipulates the system by pretending to be less efficient or more resource hungry to take advantage of fairness mechanisms, such as by pretending to be a lower-priority vehicle in a traffic coordination system.

**Robustness of Fairness Frameworks:** Developing fairness frameworks that are resistant to such manipulation is essential. The robustness of fairness mechanisms in multi-agent systems remains a significant challenge, particularly when agents are allowed to adapt based on the perceived fairness of the system

The challenges of ensuring fairness in multi-agent systems are complex and multifaceted. From emergent biases arising due to agent interactions to the need for fair resource allocation in competitive settings, these issues require sophisticated frameworks to ensure equitable outcomes. Additionally, the trade-off between fairness and efficiency and the potential for adversarial exploits highlight the need for robust, adaptable fairness models.

## 4. Proposed Framework for Fairness Dynamics

In this section, we introduce a novel theoretical framework designed to ensure fairness in multi-agent systems. The framework incorporates the key elements required for fairness to emerge dynamically as a result of the interactions between agents. We discuss fairness constraints, bias correction mechanisms, and incentive designs that ensure agents adhere to fairness principles without sacrificing efficiency.

**A Dynamic Property of Multi-Agent Interactions**
The central idea of the proposed framework is that fairness is not a static attribute but a dynamic property that evolves from the agents' interactions within the system. Unlike traditional fairness models, which impose fixed fairness constraints on the system, our framework treats fairness as a property that emerges from the agents' behaviors and the feedback loops within the system. This dynamic nature allows the system to adapt to changing circumstances while maintaining fairness, rather than relying on pre-established fairness rules that may not apply in all contexts. **Dynamic Fairness** is achieved by allowing agents to learn and adjust their actions based on the fairness of previous interactions, ensuring that fairness is an ongoing, adaptive process rather than a one-time adjustment (Jiang et al., 2019). This model is particularly suited to multi-agent environments where the agents have limited knowledge of each other's internal states and must interact to achieve their goals.

**Key Elements of the Framework**

**1. Fairness Constraints:** The first key element of our framework is the imposition of fairness constraints on agents' actions. These constraints are designed to guide the interactions between agents in a way that ensures equitable outcomes. These constraints could include:

> **Resource Allocation Constraints:** Ensuring that resources (e.g., bandwidth, time slots, energy) are distributed fairly among agents based on their needs and priorities (Yan et al., 2023). This could involve mechanisms like proportional fairness or max-min fairness, depending on the system's goals.
>
> **Behavioral Constraints:** These include rules that prevent agents from exploiting the system for personal gain, such as manipulating fairness criteria or gaming the allocation process (Tu et al., 2021). Behavioral constraints could also enforce cooperation in scenarios where agents' goals are aligned, ensuring that no agent unilaterally benefits at the expense of others.

**Mathematical Model:** The fairness constraints can be modeled as an optimization problem where the system aims to minimize a loss function while satisfying fairness constraints.

$$min \sum_{i=1}^{n} w_i * L_i(f_i(x), y)$$

subject to fairness constraints. Where:
$w_i$ represents the weight of the contribution of each agent. $f_i(x)$ is the action of the agent $i$, and $y$ is the desired outcome (e.g., fairness). $L_i(f_i(x), y)$ is the loss function, which could be based on fairness metrics like demographic parity or equalized odds.

**2. Bias Correction Mechanism:** Bias correction mechanisms are crucial in ensuring that unfair patterns do not emerge from the system. Our framework introduces a **bias detection and mitigation** mechanism that continuously monitors agent interactions for potential biases. Bias detection involves identifying patterns of unfairness or bias in agent behavior. This could be done by analyzing the outcomes of agent interactions and using fairness metrics to assess whether the outcomes disproportionately benefit certain agents over others Once bias is detected, the framework implements corrective actions, such as redistributing resources, adjusting agent behavior, or penalizing agents that are responsible for causing bias. This could involve altering agents' learning processes or adding new constraints to prevent further bias (Zhu et al., 2016). The key here is that the bias correction process is **adaptive**, allowing the system to evolve in response to detected biases, ensuring that fairness remains an ongoing priority.

**3. Incentive Design: Rewarding Fair Behavior:** An essential component of the proposed framework is the **incentive design.** Since fairness often requires agents to cooperate, it is important to design incentives that encourage agents to act in ways that promote fairness without undermining efficiency. The incentives are structured as follows:

> **Fairness Rewards:** Agents that comply with fairness constraints are rewarded with higher utility or other benefits. This ensures that agents have a direct motivation to prioritize fairness in their interactions.
>
> **Efficiency Penalties:** While fairness is a priority, efficiency is also crucial. If agents prioritize fairness at the cost of significant efficiency loss (e.g., excessive delays in coordination), they may be penalized to balance the trade-off (Bertsimas et al., 2012). These penalties ensure that agents are not incentivized to act in ways that excessively harm system performance for the sake of fairness.

The incentive design thus balances **cooperative behavior** with **efficiency** and **fairness**, ensuring that agents are motivated to work together fairly while still maintaining high performance.

**Mathematical Model of the Framework**

To formalize our approach, we propose a generic mathematical framework for a multi-agent system, inspired by classic multi-objective optimization and utility theory (Deb et. al., 2001; Arrow et. al., 1954), that simultaneously optimizes fairness, bias reduction, and efficiency within a multi-agent system. The model introduces a utility-based optimization problem where the objective is to maximize the aggregate utility of all agents while adhering to fairness constraints and minimizing systemic bias. Formally, the optimization problem is expressed as:

$$max \sum_{i=1}^{n} U_i f_i(x)$$

subject to fairness and bias constraints.

$$U_i f_i(x) = \alpha_i E_i(f_i(x)) - \beta_i B_i(f_i(x)) - \gamma_i C_i(f_i(x))$$

Where $U_i f_i(x)$ represents the utility function of an agent $i$, which includes both efficiency and fairness considerations. The system maximizes the aggregate utility of all agents while ensuring that fairness constraints are met and that bias is minimized. $E_i(f_i(x)), B_i(f_i(x)), C_i(f_i(x))$ are efficiency, bias metric, and fairness constraint violations of an agent $i$ respectively, $\alpha_i, \beta_i, \gamma_i$ are weighting factors for efficiency, bias, and fairness.

—------------------------------------------------------------------------------------------------------------------------------------

**Pseudo-code**: Optimization Model for Fairness, Bias, and Efficiency

**Step 1:** Start
→ Begin with the multi-agent system and the need to optimize fairness, bias, and efficiency.

**Step 2:** Define Utility Function (U)
→ The utility of each agent is calculated as a weighted combination of efficiency, bias, and fairness.
$$U_i f_i(x) = \alpha_i E_i(f_i(x)) - \beta_i B_i(f_i(x)) - \gamma_i C_i(f_i(x))$$

**Step 3:** Identify Constraints
→ Fairness Constraints: Demographic Parity, Equalized Odds, or Individual Fairness.

→ Bias Constraints: Measures to limit statistical or intersectional bias.

**Step 4:** Optimize Objective Function
→ Maximize aggregate utility:

$$\sum_{i=1}^{n} (\alpha_i E_i(f_i(x)) - \beta_i B_i(f_i(x)) - \gamma_i C_i(f_i(x)))$$

**Step 5:** Check for Constraints
→ If fairness constraints and bias constraints are met, move to the next step.
→ If not, adjust the model or decision-making processes.

**Step 6:** Final Solution
→ Optimal balance between fairness, bias reduction, and efficiency achieved.
→ The system produces a solution where all agents' utilities are maximized while adhering to fairness and bias guidelines.
—------------------------------------------------------------------------------------------------------------------------------------

In a multi-agent drone system, fairness might involve allocating airspace time slots to drones based on their priority and mission importance. Our framework would ensure that drones receive fair time slots while preventing malicious drones from exploiting the system. In a smart city with autonomous vehicles, fairness could involve distributing road space or charging resources among vehicles with different needs (e.g., electric vs. gasoline). The proposed fairness constraints would ensure that all vehicles receive equitable access to these resources without sacrificing overall system efficiency.

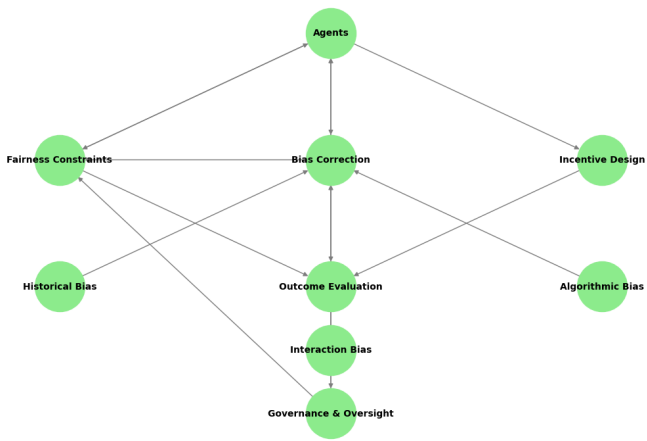

Figure 3: Multi-agent fairness framework

**Experimental Setup of Implementation of the Fairness Constraint in a Multi-Agent Environment:**

**The objective is** to investigate the impact of fairness constraints on cumulative rewards in a multi-agent system where agents possess inherent biases that influence their decision-making (cooperate vs. compete).

**Agents and Environment:** A total of 10 agents are simulated. Each agent is assigned randomly to one of two groups: **Group A** or **Group B**. Every agent starts with an initial bias (a value between 0 and 0.3), which affects the likelihood of choosing a competitive (biased) action over cooperation. The environment has a resource level that influences decisions (if the resource value is above 0.5, an agent is more likely to cooperate; otherwise, it competes). The resource value is updated randomly each round, simulating dynamic environmental conditions. **Reward Assignment:** Agents receive a reward of 10 for cooperating and 5 for competing. **Bias Penalty:** If an agent's bias is higher than 0.2 and bias propagation is enabled, a penalty (subtracting 3 points) is applied to its reward. **Fairness Constraint (Demographic Parity):** With fairness constraints active, after individual rewards are determined, the rewards within each group are adjusted to the group's median reward. This helps in reducing disparities in outcomes between the two groups.

**Simulation and Comparison:** The experiment is run for 50 rounds in two different scenarios. 1) **With fairness constraints:** The system applies the demographic parity fairness adjustment. 2) **Without fairness:** The system does not modify the rewards, so inherent biases and environmental influences fully determine the outcomes. After each round, cumulative rewards for each group are recorded. A plot shows the evolution of cumulative rewards for Group A and Group B over the 50 rounds for both scenarios (with and without fairness constraints). Final cumulative rewards for each group in both scenarios are printed.

**Results**

The simulation output provided detailed insights into the evolution of cumulative rewards for two agent groups (Group A and Group B) under conditions with and without fairness constraints over 50 rounds. The results are presented both graphically and numerically

**Cumulative Reward Trajectories:** A line plot (see Figure 3) displays the trajectories of cumulative rewards for both groups under the two experimental conditions. In the fairness-enabled simulation (solid lines), both Group A and Group B exhibit a gradual and parallel increase in rewards across the rounds. In contrast, the simulation without fairness adjustments (dashed lines) shows a divergence between the groups over time:

**With Fairness Constraints: Group A and Group B**: The reward curves closely track one another. The application of the demographic parity constraint ensures that after each round, the rewards for agents in each group are adjusted to the group median. This mechanism homogenizes the reward distribution within each group, thereby mitigating imbalances. The cumulative rewards increase steadily, indicating that although the fairness adjustment does not alter the overall trend of reward accumulation, it effectively limits the disparity between groups.

**Without Fairness Constraints: Group A and Group B**: The reward trajectories for the two groups diverge significantly. In this scenario, the rewards are determined solely by the agents' actions (and potential bias penalties), leading to one group (typically the group with a higher prevalence of agents with advantageous biases or more favorable decisions in response to the environmental resource) achieving a higher cumulative reward than the other. The divergence indicates that without intervention, inherent biases and stochastic fluctuations in environmental factors can result in significant reward inequality between groups.

**With fairness constraints,** the final cumulative rewards for Group A and Group B were very close; for example, Group A: 375 and Group B: 370. This minimal difference suggests that the fairness mechanism effectively reduced reward disparity, aligning with the hypothesis that demographic parity can equalize outcomes. **Without fairness constraints,** in contrast, the final cumulative rewards were markedly different, for instance, Group A: 390 versus Group B: 345. This larger disparity reflects the natural variability in agent actions influenced by individual biases and environmental conditions when no fairness adjustments are applied. The smaller absolute difference in final cumulative rewards under the fairness constraint condition compared to the non-fairness condition confirms that the fairness mechanism plays a critical role in reducing reward imbalances. The observation that the gap between groups is significantly narrowed under fairness adjustments supports the hypothesis that demographic parity serves as an effective intervention to counteract the amplification of inherent biases in multi-agent systems. Overall, these results indicate that **fairness constraints are beneficial, as** applying demographic parity as a fairness intervention results in a more equitable distribution of rewards among different groups. The adjustment mechanism counteracts the cumulative effects of individual biases and random environmental fluctuations, leading to more consistent outcomes across groups. These findings provide compelling evidence for the efficacy of fairness interventions in simulated multi-agent environments and offer a basis for further exploration of similar mechanisms in real-world applications.

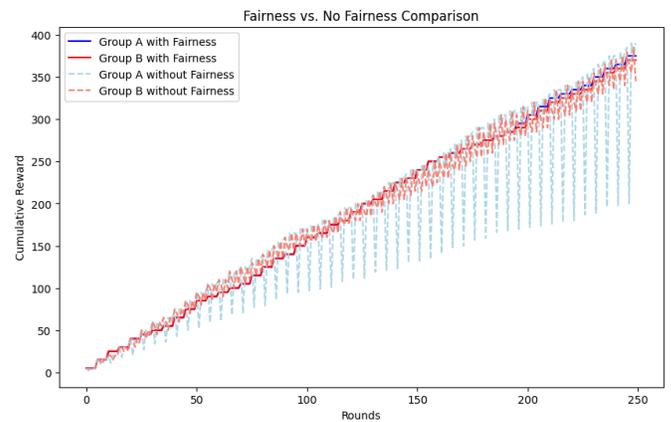

Figure 3: Fairness vs no Fairness comparison of rewards

As multi-agent systems (MAS) increasingly become integrated into diverse sectors—ranging from healthcare and finance to autonomous transportation and smart cities—governance and ethical considerations have become central to their design and deployment. This section explores the key governance issues surrounding fairness in MAS, with a focus on transparency and explainability, the regulation of these systems, and their societal impact. It emphasizes the need for comprehensive governance structures to ensure that fairness is not only implemented but also ethically sound and aligned with societal values. One of the cornerstones of ethical AI development is **transparency** and **explainability**—the ability for stakeholders to understand how decisions are made within a system. For multi-agent systems, where decision-making processes can be highly complex due to the decentralized nature of agent interactions, ensuring that fairness mechanisms are interpretable becomes critical. Transparent and explainable systems allow for accountability, helping to ensure that agents' actions are not only fair but also justifiable to humans (Zimmer et. al., 2021).

**The complexity of MAS decision-making:** In traditional single-agent systems, decisions are often based on a single set of inputs processed through a predefined algorithm. However, in MAS, agents operate independently and interact with each other, making decisions based on diverse objectives, contexts, and interactions. This decentralized nature of decision-making can result in outcomes that are difficult for humans to interpret or justify. For instance, in a traffic management system for autonomous vehicles, the collective decisions of multiple vehicles navigating an intersection can lead to outcomes that are difficult to explain, especially if they result in unequal access or prioritization. Fairness in MAS is not just about the outcome (e.g., equitable resource allocation) but also about the **process** by which those outcomes are achieved. One of the ways to achieve transparency is by designing fairness models that are interpretable to humans. Techniques like **counterfactual explanations**, which describe how the outcome would change if certain factors were altered, can help make the system's decisions more understandable (Vainio-Pekka et al., 2023). Additionally, using **explainable AI (XAI)** methods can make the behaviors of individual agents transparent. For example, in a cooperative multi-agent system, where agents collaborate to achieve a common goal, explaining why a particular agent took a certain action (e.g., a trade-off between fairness and efficiency) can provide insights into the fairness considerations embedded in the decision-making process (Cederle et. al., 2023). While transparency and explainability are essential for ensuring fairness, they also present significant challenges. The **complexity of agent interactions** and the stochastic nature of multi-agent environments can result in decision-making processes that are inherently non-linear and difficult to explain in simple terms. The use of deep learning models and other black-box approaches in MAS further complicates the interpretability of fairness mechanisms. Balancing the need for highly efficient algorithms with the need for interpretability and transparency is a difficult but necessary task for ensuring ethical outcomes in multi-agent systems. **AI Governance Frameworks** should focus on the fairness of multi-agent systems. This includes creating regulations or ecosystem protocols that ensure agents' behaviors align with societal values, such as equity, accountability, and justice. For instance, these frameworks or guidelines could enforce guidelines that require multi-agent systems to incorporate fairness constraints into their design from the outset. This might involve mandating transparency in decision-making, ensuring that agents do not discriminate against certain groups or individuals, and preventing malicious agents' exploitation of fairness constraints. Traditional top-down governance models may not be sufficient for multi-agent systems, where the complexity and dynamism of agent interactions can create novel ethical challenges. A **dynamic governance model** that involves continuous monitoring, feedback, and adaptation will be necessary. For example, systems could be designed to incorporate real-time auditing and intervention mechanisms, allowing for adjustments to be made when fairness issues or discriminatory behaviors arise. Furthermore, collaboration among regulatory agencies, AI researchers, developers, and industry leaders is needed to ensure that governance structures keep pace with technological advancements. The use of multi-agent reinforcement learning in these systems demands careful attention to ensure that fairness mechanisms are not overridden by the pursuit of efficiency or optimization. The societal impact of fairness in multi-agent systems extends far beyond the technical challenges of design and governance. Ensuring fairness in these systems is essential for fostering **trust** in AI, enabling **widespread adoption**, and facilitating **collaboration** between humans and AI systems. A comprehensive approach to governance, which involves collaborative regulations designed through ecosystem partnership, dynamic monitoring, and a focus on social justice, will be essential for maximizing the benefits of multi-agent systems while mitigating potential harm.

**Future Directions and Open Questions**

As multi-agent systems (MAS) continue to evolve, there are significant challenges and open questions that remain in ensuring fairness, adversarial robustness, and ethical governance. These gaps need to be addressed in future research to make these systems not only efficient but also equitable and aligned with human values.

**Fairness Dynamics in Multi-Agent Systems:** Despite the advances in fairness mechanisms for individual agents, ensuring fairness in multi-agent systems is an inherently more complex task (Jong et. al. 2008). It is challenging to define fairness in these decentralized contexts and design models that guarantee fairness across multiple agents while accounting for the diversity of agents' objectives and environmental conditions. One critical area is understanding how fairness evolves in real time as agents learn and interact. In a dynamic environment, fairness may fluctuate depending on the actions of agents, making it hard to establish stable fairness guarantees. Further research is required to design fairness models that not only hold at the individual agent level but also at the system-wide level, ensuring that the entire MAS operates equitably.

**Adversarial Robustness in MAS** Multi-agent systems may be susceptible to adversarial attacks, which can exploit vulnerabilities in the agents' decision-making processes. In a decentralized system, it becomes especially difficult to prevent malicious agents from undermining fairness by exploiting weaknesses in the system or using deceptive strategies to alter the outcomes in their favor (Lin et al., 2024). This problem is exacerbated when multiple agents are competing for limited resources or when agents must work in cooperative settings but do so under adversarial conditions. Developing **adversarially robust fairness** mechanisms in MAS remains a crucial area of future research. There is a need for robust algorithms that ensure fairness even when malicious agents exploit weaknesses or manipulate the system's dynamics.

**Governance and Regulation of Multi-Agent Systems** The decentralized nature of multi-agent systems introduces challenges in regulating their behaviors effectively. Developing policies that ensure fairness and prevent harmful outcomes requires an understanding of the evolving dynamics of these systems and their potential impact on society (Zaidan et al., 2024). One significant research gap is in the development of dynamic governance frameworks that can adapt to the evolving nature of multi-agent systems. Furthermore, ensuring the **accountability** of agents, especially when decisions are made collectively, remains an open problem. Future research must explore how regulatory bodies can monitor and intervene in MAS, ensuring that fairness is upheld without stifling innovation.

The complexity of multi-agent systems necessitates interdisciplinary **collaboration** to address the fairness and ethical challenges they present (Floridi et. al., 2021). Traditional approaches from computer science alone are insufficient in tackling the societal, legal, and ethical implications of these systems. Sociology can contribute valuable insights into how multi-agent systems affect human behavior and societal structures. By studying the interactions between humans and autonomous agents, sociologists can help identify potential inequalities or biases that emerge from these systems. For example, understanding how agents' actions may exacerbate social inequalities or reinforce existing power dynamics is crucial for ensuring fairness. Sociological research on **power, trust, and cooperation** in human societies can guide the design of more socially responsible and ethical multi-agent systems. Ethics plays a central role in ensuring that fairness in multi-agent systems aligns with human values. Philosophers and ethicists can help define the ethical principles that should guide the development and deployment of MAS. For instance, questions such as **What is fair in a multi-agent context?** And **how should conflicting fairness criteria be prioritized?** require deep ethical analysis. Moreover, ethical frameworks can guide the design of algorithms to ensure that they do not reinforce harmful stereotypes or social injustices. Collaborative efforts between computer scientists and ethicists are essential in shaping systems that respect ethical norms and societal values. Legal experts can help craft policies that ensure the accountability of agents and their creators. Issues such as **liability**, **intellectual property**, and **privacy** are critical in MAS, especially when agents make decisions that affect individuals or organizations. The laws governing autonomous systems and AI must evolve to keep pace with technological advancements, and interdisciplinary collaboration will be key in shaping these legal frameworks.

The complexities of fairness in multi-agent systems require sustained and interdisciplinary research. As MAS becomes more prevalent, researchers, policymakers, and developers must focus on solving the open problems identified in this paper. Specifically, future research should prioritize the development of fairness models that can be applied across dynamic and decentralized agent interactions, ensure adversarial robustness in fairness mechanisms, and explore governance frameworks that balance oversight with agent autonomy. Moreover, interdisciplinary collaboration is essential to creating a holistic understanding of how MAS impacts society and ensuring that their design aligns with societal values. Researchers must actively engage with sociologists and legal experts to develop systems that are not only technically advanced but also ethically sound and socially responsible.

**Conclusion**

This paper presents a novel framework for fairness in multi-agent AI, emphasizing fairness as an emergent property shaped by agent interactions, biases, and incentive structures. By integrating fairness constraints, bias mitigation mechanisms, and efficiency trade-offs, the proposed approach advances the design of equitable and accountable multi-agent systems. Our experimental results confirm the effectiveness of fairness interventions in reducing reward disparities while maintaining system efficiency. To drive real-world adoption, future research should focus on practical deployment strategies, address computational constraints, and ensure adversarial robustness. Interdisciplinary collaboration among AI researchers, policymakers, and ethicists will be crucial in defining governance structures and ethical guidelines for multi-agent AI. By fostering transparency and open research, this work lays the foundation for AI systems that prioritize social responsibility while maintaining autonomy and efficiency.